\documentstyle[twoside,fleqn,espcrc2,epsf]{article}
\newcommand{\nn}{\nonumber}
\newcommand{\be}{\begin{equation}}
\newcommand{\ee}{\end{equation}}
\newcommand{\bea}{\begin{eqnarray}}
\newcommand{\eea}{\end{eqnarray}}

\def\bfnabla{\mbox{\boldmath $\nabla$}}
\def\bfSigma{\mbox{\boldmath $\Sigma$}}
\def\bfsigma{\mbox{\boldmath $\sigma$}}
\def\bfPi{\mbox{\boldmath $\Pi$}}
\def\lQ{\Lambda_{\rm QCD}}
\def\al{\alpha}
\def\als{\alpha_{\rm s}}
\def\siml{{\ \lower-1.2pt\vbox{\hbox{\rlap{$<$}\lower6pt\vbox{\hbox{$\sim$}}}}\ }} 

\def\lla{\langle\!\langle}
\def\rra{\rangle\!\rangle}
\newcommand{\Appendix}[1]%
    {%
     \section{#1}%
      }

\begin{document}
\def\siml{{\ \lower-1.2pt\vbox{\hbox{\rlap{$<$}\lower6pt\vbox{\hbox{$\sim$}}}}\ }} 
\def\bfnabla{\mbox{\boldmath $\nabla$}}
\def\bfSigma{\mbox{\boldmath $\Sigma$}}
\def\bfsigma{\mbox{\boldmath $\sigma$}}
\def\als{\alpha_{\rm s}}
\def\al{\alpha}
\def\lQ{\Lambda_{\rm QCD}}
\def\vs{V^{(0)}_s}
\def\vo{V^{(0)}_o}
\newcommand{\ttbs}{\char'134}
\newcommand{\AmS}{{\protect\the\textfont2 A\kern-.1667em\lower.5ex\hbox{M}\kern-.125emS}}

\title{The non-perturbative QCD potential}

\author{Antonio Vairo\address{
Institut f\"ur Theoretische Physik, Universit\"at Heidelberg\\ 
Philosophenweg 16, D-69120 Heidelberg, FRG}
\thanks{Alexander von Humboldt fellow}}

\begin{abstract}
The potential between heavy quarks can be rigorously defined 
in a QCD effective field theories framework. I discuss the general 
situation when the coupling constant may not be considered small and 
provide an explicit expression for the potential valid up to order $1/m^2$. 
Spin dependent and spin independent terms are expressed in terms 
of Wilson loops, completing an ideal journey started 
twenty years ago with the classical work of Eichten and Feinberg.
\end{abstract}

\maketitle

\section{INTRODUCTION}
The mass of the $b$ and the $c$ quarks is usually considered large enough 
to treat as non-relativistic (NR) heavy-quark--antiquark systems made up by them (heavy quarkonia: 
$\psi$, $\Upsilon$, $B_c$, ...). Therefore, these systems are  
characterized by, at least, three separated scales: hard (the mass $m$ of the heavy
quarks), soft (the momentum scale $|{\bf p}| \sim mv$, $ v \ll 1$), 
and ultrasoft (the typical kinetic energy $E \sim mv^2$
of the heavy quark in the bound-state system). This NR picture is 
supported by the success of traditional potential models. 
In this case a description of heavy quarkonia systems in terms of a NR  
Schr\"odinger equation is assumed and the interaction potential is shaped 
in order to reproduce as best as possible the data (for some reviews see \cite{rev0,rev1,gunnar}). 
The fact that such a description turns out to work reasonably well 
even with simple {\it ans\"atze} on the potential (like, for instance, the 
Cornell potential) suggests that a NR quantum-mechanical description 
of heavy quarkonium systems, may, indeed, be appropriate and justified from QCD.  

\begin{figure}[t]
\vskip -2.2truecm
\makebox[0.0truecm]{\phantom b}
\put(0,0){\epsfxsize=8truecm \epsffile{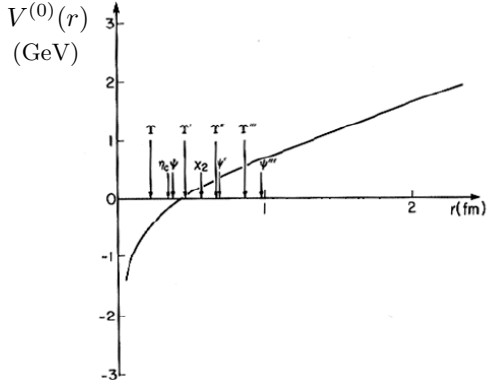}}
\put(15,210){$V^{(0)}(r)$}
\put(17,197){\small (GeV)}
\vskip -33truemm
\caption{\it The size of some heavy quarkonia is shown with respect to the Cornell potential \cite{gois}.} 
\vspace{-10mm}
\label{plpot}
\end{figure}

Fig. \ref{plpot}, taken from \cite{gois}, shows the static potential 
versus the size of different heavy quarkonium systems. It suggests that heavy quarkonium 
ground states may belong to the region of validity of perturbative QCD (the region where the static 
potential has a $1/r$ behaviour). The perturbative QCD potential has been calculated 
up to order $\als^4$ (with respect to the energy level) in \cite{Peter} (the two-loop static potential) 
and in \cite{Gupta} (the one-loop $1/m$ potential and the tree-level $1/m^2$ potentials). 
Non-potential effects show up in perturbation theory at the NNNLO and have been investigated in 
\cite{pnrqcd,pnrqcd1}. Fig. \ref{plpot} shows, however, that most of the quarkonia states 
lie in a region where the inverse of the size of the system is close to the scale $\lQ$ 
of non-perturbative physics (i.e. in the region where the static potential 
starts to rise linearly in $r$). In this situation the potential can no longer be expressed 
as an expansion in $\als$, but, starting from the seminal work of Wilson \cite{wilson}, 
it has been related to Wilson loop operators. These operators can be eventually 
calculated on the lattice \cite{latpot} or in QCD vacuum models \cite{mod}. 

Despite the fact that the non-perturbative QCD potential has been investigated 
for more than twenty years \cite{Brown,spin1,spin2,BMP,cor1,thesis,chen,BV1} 
only recently the complete (in pure gluodynamics) and correct expression valid up to order $1/m^2$ 
has been calculated \cite{m1,m2}. Here, I will report about these last progresses.
The theoretical framework is NRQCD \cite{NRQCD} and pNRQCD \cite{pnrqcd0,pnrqcd},
which are the suitable effective field theories for systems made up by two heavy quarks.
NRQCD is obtained from QCD by integrating out the hard scale $m$.  
It is characterized by an ultraviolet cut-off much smaller than the mass $m$ and
much larger than any other scale, in particular much larger than $\lQ$.  This
means that the matching from QCD to NRQCD can be done perturbatively,
as well as within an expansion in $1/m$ \cite{Manohar,Match}. 
By integrating out the momentum scale $mv$ one is left with the effective field theory called pNRQCD,  
where the soft and ultrasoft scales have been disentangled and where the connection between NRQCD 
and a NR quantum-mechanical description of the system may be formalized in a systematic way. 
We will assume that the matching between NRQCD and pNRQCD can be performed 
order by order in the $1/m$ expansion. 

In \cite{m1,m2} the matching of NRQCD to pNRQCD has been performed at order $1/m^2$ 
in the general situation $\lQ \siml mv$. This has been proved 
to be equivalent to compute the heavy quarkonium potential at order $1/m^2$. 
More precisely, a pure potential picture emerges at the leading order in the ultrasoft 
expansion under the condition that all the gluonic excitations have a gap of $O(\lQ)$.
Higher order effects in the $1/m$ expansion as well as extra ultrasoft degrees of freedom 
such as hybrids and pions can be systematically included and may eventually affect the leading 
potential picture (like in the perturbative regime ultrasoft gluons \cite{pnrqcd}).
The obtained expression for the potential is also correct at any power in $\als$
in the perturbative regime. In the following I will introduce the formalism and outline 
the derivation of the quarkonium potential in pNRQCD. I will discuss the results and briefly 
the non-perturbative power counting.

\section{NRQCD}
\label{secnrqcd}
After integrating out the hard scale $m$, we obtain NRQCD \cite{NRQCD}. 
Neglecting operators involving light quark fields \cite{ManBauer}, the NRQCD Hamiltonian 
for a quark of mass $m_1$ and an antiquark of mass $m_2$ up to $O(1/m^2)$ is given by:
\begin{eqnarray}
&&\!\!\!\!\!\!\!\!\!\!
H = H^{(0)}+ \sum_{\ell+n=1,2} {H^{(\ell,n)} \over m_1^\ell m_2^n}, 
\label{HH}\\
&&\!\!\!\!\!\!\!\!\!\!
H^{(0)} = \int d^3{\bf x} {1\over 2}\left( {\bfPi}^a{\bfPi}^a +{\bf B}^a{\bf B}^a \right),
\label{H0}\\
&&\!\!\!\!\!\!\!\!\!\!
H^{(1,0)} = - {1 \over 2} \int d^3{\bf x} \, \psi^\dagger \left( {\bf D}^2 
+ g c_F^{(1)} \bfsigma \cdot {\bf B}\right) \psi,  
\label{H1}\\
&&\!\!\!\!\!\!\!\!\!\! 
H^{(0,1)} = - H^{(1,0)}(\psi \leftrightarrow \chi; 1 \leftrightarrow 2)
\label{H01}\\ 
&&\!\!\!\!\!\!\!\!\!\!
H^{(2,0)} = \int d^3{\bf x} \,
\psi^\dagger\Biggl\{ - c_D^{(1)\,\prime} \, g { \left[{\bf D} \cdot, {\bf E}  \right] \over 8 }\nn \\
&&\qquad\qquad 
- i c_S^{(1)} \, g {  \bfsigma \cdot
\left[{\bf D} \times, {\bf E} \right] \over 8 } \Biggr\} \psi  \nn \\
&&\qquad 
- \int d^3{\bf x} \, d_3^{(1)\,\prime} g f_{abc}G^a_{\mu\nu} G^b_{\mu\al} G^c_{\nu\al},
\label{H20}\\
&&\!\!\!\!\!\!\!\!\!\!
H^{(0,2)} = H^{(2,0)}(\psi \leftrightarrow \chi; 1 \leftrightarrow 2),
\label{H02}\\
&&\!\!\!\!\!\!\!\!\!\! 
H^{(1,1)} = - \! \int \!\! d^3{\bf x} \left( d_{ss}  \psi^{\dag} \psi \chi^{\dag} \chi
+  d_{sv}  \psi^{\dag} {\bfsigma} \psi \chi^{\dag} {\bfsigma} \chi \right.\nn\\
&&\!\!\!\!\!\!\!\!\!\!
\left. +  d_{vs}  \psi^{\dag} {\rm T}^a \psi \chi^{\dag} {\rm T}^a \chi
+  d_{vv}  \psi^{\dag} {\rm T}^a {\bfsigma} \psi \chi^{\dag} {\rm T}^a {\bfsigma} \chi \right), 
\label{H11}
\end{eqnarray}
where $\psi$ is the Pauli spinor field that annihilates the fermion 
and $\chi$ is the Pauli spinor field that creates the antifermion, $i{\bf D}=i\bfnabla+g{\bf A}$,
$[{\bf D \cdot, E}]$ $=$ ${\bf D \cdot E} - {\bf E \cdot D}$ and 
$[{\bf D \times, E}]$ $=$ ${\bf D \times E -E \times D}$.
$\bfPi^a$ $=$ ${\bf E}^a + O(1/m^2)$ is the canonical momentum conjugated to ${\bf A}^a$ 
and the physical states are constrained to satisfy the Gauss law:
${\bf D}\cdot {\bfPi}^a \vert {\rm phys} \rangle$ $=$  
$g (\psi^\dagger T^a \psi$ $+$  $\chi^\dagger T^a \chi) \vert {\rm phys} \rangle$.
The coefficients $c_F$, $c_D^\prime$, $c_S$, $d_2$ and $d_3^\prime$ 
can be found in Ref. \cite{Manohar,m2} and $d_{ij}$ ($i,j=s,v$) in
\cite{Match} for the $\overline{MS}$ scheme.

In the static limit the one-quark--one-antiquark sector of the Fock space may be spanned by
$\vert \underbar{n}; {\bf x}_1 ,{\bf x}_2  \rangle^{(0)}$ 
$=$  $\psi^{\dagger}({\bf x}_1) \chi_c^{\dagger} ({\bf x}_2) |n;{\bf x}_1 ,{\bf x}_2\rangle^{(0)}$, 
where $|\underbar{n}; {\bf x}_1 ,{\bf x}_2\rangle^{(0)} $ is a gauge-invariant
eigenstate (up to a phase) of $ H^{(0)}$, as a consequence of the Gauss law, 
with energy $E_{n}^{(0)}({\bf x}_1 ,{\bf x}_2)$, and  $\chi_c ({\bf x})=i\sigma^2 \chi^{*} ({\bf x})$.  
$|n;{\bf x}_1 ,{\bf x}_2\rangle^{(0)}$ encodes the gluonic content of the
state, i.e.  it is annihilated by $\chi_c({\bf x})$ and $\psi ({\bf x})$ ($\forall {\bf x}$). 
The normalization condition is 
$^{(0)}\langle \underbar{m}; {\bf x}_1 ,{\bf x}_2|\underbar{n}; {\bf y}_1 ,{\bf y}_2\rangle^{(0)}$  
$=$ $\delta_{nm} \displaystyle \prod_{j=1}^2 \delta^{(3)} ({\bf x}_j -{\bf y}_j)$.
The positions ${\bf x}_1$ and ${\bf x}_2$ of the quark and antiquark respectively 
are good quantum numbers for the static solution $|\underbar{n};{\bf x}_1 ,{\bf x}_2 \rangle^{(0)}$; 
$n$ generically denotes the remaining quantum numbers, which are classified by the irreducible 
representations of the symmetry group $D_{\infty h}$ (substituting the parity
generator by CP). We also choose the basis such that $T|\underbar{n};{\bf x}_1
,{\bf x}_2 \rangle^{(0)}= |\underbar{n};{\bf x}_1
,{\bf x}_2 \rangle^{(0)}$, where $T$ is the time-inversion operator. 
The ground-state energy $E_0^{(0)}({\bf x}_1,{\bf x}_2)$ can be associated to the static potential 
of the heavy quarkonium under some circumstances (see Sec. \ref{secpnrqcd}). 
The remaining energies $E_n^{(0)}({\bf x}_1,{\bf x}_2)$, $n\not=0$, are usually associated 
to the potentials describing heavy hybrids or heavy quarkonium (or other heavy hybrids) plus glueballs. 
They can be computed on the lattice (see, for instance, \cite{michael} and \cite{nora99}). 
Translational invariance implies that  $E_n^{(0)}({\bf x}_1,{\bf x}_2) = 
E_n^{(0)}(r)$, where ${\bf r}={\bf x}_1-{\bf x}_2$. 

Beyond the static limit, but still working order by order in $1/m$,
the normalized eigenstates, $|\underbar{n}; {\bf x}_1 ,{\bf x}_2\rangle$,
and eigenvalues, $E_n({\bf x}_1 ,{\bf x}_2; {\bf p}_1, {\bf p}_2)$, 
of the Hamiltonian $H$ satisfy the equations \vspace{-5mm}\\
\bea 
&&\!\!\!\!\!\!\!\!\!\! 
H |\underbar{n}; {\bf x}_1 ,{\bf x}_2\rangle = \int d^3x_1^\prime d^3x_2^\prime 
|\underbar{n}; {\bf x}_1^\prime ,{\bf x}_2^\prime \rangle \nn\\
&& \times 
E_n({\bf x}_1^\prime,{\bf x}_2^\prime; {\bf p}_1^\prime, {\bf p}_2^\prime)
\prod_{j=1}^2 \delta^{(3)} ({\bf x}_j^\prime -{\bf x}_j), 
\label{bornschroe} \\
&&\!\!\!\!\!\!\!\!\!\! 
\langle \underbar{m}; {\bf x}_1 ,{\bf x}_2|\underbar{n}; {\bf y}_1 ,{\bf y}_2\rangle = 
\delta_{nm} \prod_{j=1}^2 \delta^{(3)} ({\bf x}_j -{\bf y}_j).
\label{bornnorm}
\vspace{-6mm}
\eea
Note that the positions ${\bf x}_1$ and ${\bf x}_2$ of the static solution still label 
the states even if the position operator does not commute with $H$ 
beyond the static limit. We are interested in the eigenvalues $E_n$. 
$E_0$ corresponds to the quantum-mechanical Hamiltonian of the heavy quarkonium (in some specific 
situation). The other energies $E_n$ for $n \!> \!0$ are related 
to the quantum-mechanical Hamiltonian of  higher gluonic excitations between heavy quarks. 
Expanding Eqs. (\ref{bornschroe}) and (\ref{bornnorm}) around the static solution 
we get up to $O(1/m^2)$ \vspace{-4mm}\\ 
\bea
&&\!\!\!\!\!\!\!\!\!\! 
E_n({\bf x}_1,{\bf x}_2; {\bf p}_1, {\bf p}_2)
\prod_{j=1}^2 \delta^{(3)} ({\bf x}_j^\prime -{\bf x}_j) =  \nn\\
&&\!\!\!\!\!\!\!\!\!\! 
E_n^{(0)}({\bf x}_1,{\bf x}_2) 
\prod_{j=1}^2 \delta^{(3)} ({\bf x}_j^\prime -{\bf x}_j) \nn\\
&&\!\!\!\!\!\!\!\!\!\! 
+\, ^{(0)}\langle \underbar{n}; {\bf x}_1 ,{\bf x}_2 \vert
\sum_{\ell+j=1,2} {H^{(\ell,j)} \over m_1^\ell m_2^j} 
\vert \underbar{n}; {\bf x}_1^\prime ,{\bf x}_2^\prime \rangle^{(0)} 
\nn \\
&&\!\!\!\!\!\!\!\!\!\! 
- \, {1\over 2}\sum_{k\neq n} \int d^3y_1 \, d^3y_2 \,\nn\\
&&\qquad 
\times  ^{(0)}\langle \underbar{n}; {\bf x}_1 ,{\bf x}_2 \vert 
\sum_{\ell+j=1} {H^{(\ell,j)} \over m_1^\ell m_2^j} 
\vert \underbar{k}; {\bf y}_1 ,{\bf y}_2\rangle^{(0)}\,\nn\\
&&\qquad
\times ^{(0)}\langle \underbar{k}; {\bf y}_1,{\bf y}_2 \vert 
\sum_{\ell+j=1} {H^{(\ell,j)} \over m_1^\ell m_2^j} 
\vert \underbar{n}; {\bf x}_1^\prime,{\bf x}_2^\prime \rangle^{(0)} \nn\\
&&\qquad
\times \left( {1\over E_k^{(0)}({\bf y}_1,{\bf y}_2) - E_n^{(0)}({\bf x}_1^\prime,{\bf x}_2^\prime)} 
\right.\nn\\
&& \qquad\quad 
\left. + {1\over E_k^{(0)}({\bf y}_1,{\bf y}_2) - E_n^{(0)}({\bf x}_1,{\bf x}_2)} \right).
\label{En2}
\eea  
Explicit expressions for the energies $E_n$, obtained from the above formula, can be found in \cite{m1,m2}.

\section{pNRQCD}
\label{secpnrqcd}
In the static limit, the gap between different states at fixed ${\bf r}$ 
depends on the dimensionless parameter $\lQ r$. In general there
will be a set of states $\{n_{\rm us}\}$ such that $E_{n_{\rm us}}^{(0)}(r)
\sim mv^2$ for the typical $r$ of the actual physical system. We call these
states ultrasoft. The aim of pNRQCD is to describe the behaviour of the
ultrasoft states. Therefore, in order to obtain pNRQCD all the physical degrees of freedom with energies
larger than $mv^2$ are integrated out from NRQCD. In this context one may work order by order in $1/m$ (in
particular for the kinetic energy), and the calculation of the previous
section becomes the matching calculation between NRQCD and pNRQCD.
 
In the perturbative situation, $\lQ r \ll 1$, $\{n_{\rm us}\}$ corresponds to 
a heavy-quark--antiquark state, in either a singlet or an octet configuration, plus gluons and light
fermions, all of them with energies of $O(mv^2)$ \cite{pnrqcd}.  In a non-perturbative
situation, $\lQ r \sim 1$, it is not obvious  what $\{n_{\rm us}\}$ is. One may think of different 
possibilities. In particular, one could consider the situation where, because of a mass gap in QCD, the
energy splitting between the ground state and the first gluonic excitation is
larger than $mv^2$, and, because of chiral symmetry breaking of QCD, Goldstone
bosons (pions/kaons) appear. Hence, in this situation, $\{n_{\rm us}\}$ would
be the ultrasoft excitations about the static ground state, which we will call the singlet,
plus the Goldstone bosons. If one switches off the light fermions (pure
gluodynamics), only the singlet survives and pNRQCD reduces to a pure two-particle 
NR quantum-mechanical system. Therefore, the situation {\it assumed} by all potential models, 
may be  now rigorously {\it derived} under a specific set of circumstances.

Here, I shall discuss the pure singlet sector with no further reference to ultrasoft 
degrees of freedom. In this situation, pNRQCD only describes the ultrasoft excitations about the
static ground state of NRQCD. In terms of static NRQCD eigenstates, this means that only  
$|\underbar{0}; {\bf x}_1 ,{\bf x}_2\rangle^{(0)}$ is kept as an explicit degree of freedom,  
whereas $|\underbar{n}; {\bf x}_1 ,{\bf x}_2\rangle^{(0)}$ with  $n\not=0$ are integrated 
out. This provides the only dynamical degree of freedom of the theory. It is described by means 
of a bilinear colour singlet field, $S({\bf x}_1,{\bf x}_2,t)$, which has the same quantum numbers 
and transformation properties under symmetries as the static ground state of NRQCD in the 
one-quark--one-antiquark sector. In the above situation, the Lagrangian of pNRQCD reads \vspace{-5mm}\\
\be
{\cal L}_{\rm pNRQCD} = S^\dagger 
\bigg( i\partial_0 -h_s({\bf x}_1,{\bf x}_2, {\bf p}_1, {\bf p}_2)\bigg) S, 
\label{pnrqcdl}
\ee
where $h_s$ is the Hamiltonian of the singlet, ${\bf p}_1= -i \bfnabla_{{\bf x}_1}$ 
and ${\bf p}_2= -i \bfnabla_{{\bf x}_2}$. It has the following expansion up to order $1/m^2$ 
\bea
&&\!\!\!\!\!\!\!\!\!\! 
h_s ({\bf x}_1,{\bf x}_2, {\bf p}_1, {\bf p}_2) = 
{{\bf p}^2_1\over 2 m_1} +{{\bf p}^2_2\over 2 m_2} + V^{(0)}\label{hss}\\
&&\!\!\!\!\!\!\!\!\!\! 
+{V^{(1,0)} \over m_1}+{V^{(0,1)} \over m_2}+ {V^{(2,0)} \over m_1^2}
+ {V^{(0,2)}\over m_2^2}+{V^{(1,1)} \over m_1m_2}.
\nn
\eea

The integration of the higher excitations is trivial using the basis 
$|\underbar{n}; {\bf x}_1 ,{\bf x}_2\rangle$ since,
in this case, they are decoupled from $|\underbar{0}; {\bf x}_1 ,{\bf x}_2 \rangle$. 
The matching of NRQCD to pNRQCD simply consists in a renaming of things in a way such
that pNRQCD reproduces the matrix elements of NRQCD for the ground state, in
particular the energy. This fixes the matching condition \vspace{-5mm}\\
\be
E_0({\bf x}_1,{\bf x}_2, {\bf p}_1, {\bf p}_2) = h_s({\bf x}_1,{\bf x}_2, {\bf p}_1, {\bf p}_2) .
\label{singlet}
\ee

\section{$\!\!$ HEAVY $\!\!\!$ QUARKONIUM $\!\!\!$  POTENTIAL}
\label{secwilson}
To express the heavy quarkonium potential in terms of Wilson loops is quite convenient for 
lattice simulations \cite{latpot}. 
We shall use the following notations: $\langle \dots \rangle$ will stand for the average 
over the Yang--Mills action, $W_\Box$ for the rectangular static Wilson loop of dimensions 
$r\times T_W$ and $\langle\!\langle \dots \rangle\!\rangle 
\equiv \langle \dots W_\Box\rangle / \langle  W_\Box\rangle$. 
We define $\lla O_1(t_1)O_2(t_2)...O_n(t_n)\rra_c$ 
as the {\it connected} Wilson loop with $O_1(t_1)$, $O_2(t_2)$,
... $O_n(t_n)$ operators insertions for $T_W/2 \ge t_1 \ge t_2 \ge \dots \ge t_n \ge -T_W/2$. 
We also define in a short-hand notation 
$\displaystyle \lim_{T\rightarrow \infty} \equiv \lim_{T\rightarrow \infty}\lim_{T_W\rightarrow \infty}$, 
where $T\le T_W$ is the time-length appearing in the time integrals. By performing first 
the limit $T_W\rightarrow \infty$, the averages 
$\lla \dots \rra$ become independent of $T_W$ and thus invariant under global time translations. 

Using the matching condition (\ref{singlet}) and Eq. (\ref{En2}) we get in terms of Wilson 
loops \cite{m1}
\be 
V^{(0)}(r) = \lim_{T\to\infty}{i\over T} \ln \langle W_\Box \rangle, 
\label{v0}
\ee
\bea
&&\!\!\!\!\!\!\!\!\!\! 
V^{(1,0)}(r)=
-{1 \over 2} \lim_{T\rightarrow \infty}\int_0^{T}dt \, t \, \lla g{\bf E}_1(t)\cdot g{\bf E}_1(0) \rra_c 
\nn\\
&&\qquad\;  = V^{(0,1)}(r),
\label{Em12}
\eea
where ${\bf E}_j \equiv {\bf E}({\bf x}_j)$.

Let us now consider the terms of order $1/m^2$.
We define 
\begin{eqnarray*}
&&\!\!\!\!\!\!\!\!\!\! 
V^{(2,0)} = {1 \over 2}\left\{{\bf p}_1^2,V_{{\bf p}^2}^{(2,0)}(r)\right\}
+{V_{{\bf L}^2}^{(2,0)}(r)\over r^2}{\bf L}_1^2
\\ &&\!\!\!\!\!\!\!\!\!\! 
+ V_r^{(2,0)}(r) + V^{(2,0)}_{LS}(r){\bf L}_1\cdot{\bf S}_1,
\end{eqnarray*}
where ${\bf L}_j \equiv {\bf r} \times {\bf p}_j$. 
Analogous definitions held  for $V^{(0,2)}$, for which, by using invariance under charge conjugation 
plus $m_1 \leftrightarrow m_2$ transformation, we have 
$V_{{\bf p}^2}^{(2,0)}(r)$ $=$ $V_{{\bf p}^2}^{(0,2)}(r)$, 
$V_{{\bf L}^2}^{(2,0)}(r)$ $=$ $V_{{\bf L}^2}^{(0,2)}(r)$ and 
$V_{r,\, LS}^{(2,0)}(r)$ $=$ $V_{r,\, LS}^{(0,2)}(r;m_2\leftrightarrow m_1)$.
Using Eqs. (\ref{singlet}) and (\ref{En2}) we get, in terms of Wilson loops,  
\bea
&&\!\!\!\!\!\!\!\!\!\! \nn 
V_{{\bf p}^2}^{(2,0)}(r)={i \over 2}{\hat {\bf r}}^i{\hat {\bf r}}^j
\lim_{T\rightarrow \infty}\int_0^{T}dt \, t^2 
\\ &&\quad \times 
\lla g{\bf E}_1^i(t) g{\bf E}_1^j(0) \rra_c,
\label{Em20p}
\eea
\bea
&&\!\!\!\!\!\!\!\!\!\! \nn 
V_{{\bf L}^2}^{(2,0)}(r)={i \over 4}
\left(\delta^{ij}-3{\hat {\bf r}}^i{\hat {\bf r}}^j \right)
\\ &&\quad \times 
\lim_{T\rightarrow \infty}\int_0^{T}dt \, t^2 \lla g{\bf E}_1^i(t) g{\bf E}_1^j(0) \rra_c,
\label{Em20L}
\eea
\bea
&&\!\!\!\!\!\!\!\!\!\! 
V_r^{(2,0)}(r)= - {c_D^{(1)} \over 8} 
\lim_{T\rightarrow \infty}\int_0^{T}dt \, \lla [{\bf D}_1,g{\bf E}_1(t)] \rra_c 
\label{Em20} 
\\ \nn &&\!\!\!\!\!\!\!\!\!\! 
- {i c_F^{(1)\,2} \over 4}  \lim_{T\rightarrow  \infty}\int_0^{T}dt \, 
\lla g{\bf B}_1(t)\cdot g{\bf B}_1(0) \rra_c
\\ \nn &&\!\!\!\!\!\!\!\!\!\! 
-{i \over 2} \lim_{T\rightarrow \infty}\int_0^{T}dt_1\int_0^{t_1} dt_2 \int_0^{t_2}
dt_3\, (t_2-t_3)^2 
\\ \nn &&\quad \times 
\lla g{\bf E}_1(t_1)\cdot g{\bf E}_1(t_2) g{\bf E}_1(t_3)\cdot g{\bf E}_1(0) \rra_c 
\\ \nn &&\!\!\!\!\!\!\!\!\!\! 
+ {1 \over 2}\Bigg(\bfnabla_r^i \lim_{T\rightarrow \infty}\int_0^{T}dt_1\int_0^{t_1} dt_2 \, (t_1-t_2)^2
\\ \nn &&\quad \times 
\lla g{\bf E}_1^i(t_1) g{\bf E}_1(t_2)\cdot g{\bf E}_1(0) \rra_c \Bigg)
\\ \nn &&\!\!\!\!\!\!\!\!\!\! 
- {i \over 2}\left(\bfnabla_r^i V^{(0)}\right) \lim_{T\rightarrow \infty}
\int_0^{T}dt_1\int_0^{t_1} dt_2 \, (t_1-t_2)^3 
\\ \nn &&\quad \times 
\lla g{\bf E}_1^i(t_1) g{\bf E}_1(t_2)\cdot g{\bf E}_1(0) \rra_c
\\ \nn &&\!\!\!\!\!\!\!\!\!\! 
- {1 \over 2} \lim_{T\rightarrow \infty}\int_0^{T}dt_1\int_0^{t_1} dt_2 \, (t_1-t_2)^2 
\\ \nn &&\quad \times 
\lla [{\bf D}_1.,g{\bf E}_1](t_1) g{\bf E}_1(t_2)\cdot g{\bf E}_1(0) \rra_c
\\ \nn &&\!\!\!\!\!\!\!\!\!\! 
+ {i \over 8} \lim_{T\rightarrow \infty}\int_0^{T}dt \, t^2
\lla [{\bf D}_1.,g{\bf E}_1](t) [{\bf D}_1.,g{\bf E}_1](0) \rra_c
\\ \nn &&\!\!\!\!\!\!\!\!\!\! 
- {i \over 4} \Bigg( \bfnabla_r^i \lim_{T\rightarrow \infty}\int_0^{T}dt \, t^2
\lla g{\bf E}_1^i(t) [{\bf D}_1.,g{\bf E}_1](0) \rra_c \Bigg)
\\ \nn &&\!\!\!\!\!\!\!\!\!\! 
- {1 \over 4} \lim_{T\rightarrow \infty}\int_0^{T}dt \, t^3
\lla [{\bf D}_1.,g{\bf E}_1](t) g{\bf E}_1^j (0) \rra_c (\bfnabla_r^j V^{(0)})
\\ \nn &&\!\!\!\!\!\!\!\!\!\! 
+ {1 \over 4} \Bigg(\bfnabla_r^i \lim_{T\rightarrow \infty}\int_0^{T}dt \, t^3
\lla g{\bf E}_1^i(t) g{\bf E}_1^j (0) \rra_c (\bfnabla_r^j V^{(0)}) \Bigg)
\\ \nn &&\!\!\!\!\!\!\!\!\!\! 
+ {1 \over 2}(\bfnabla_r^2 V_{{\bf p}^2}^{(2,0)})
- {i \over 12} \lim_{T\rightarrow \infty}\int_0^{T}dt \, t^4
\\ \nn &&\quad \times 
\lla g{\bf E}_1^i(t) g{\bf E}_1^j (0) \rra_c (\bfnabla_r^i V^{(0)}) (\bfnabla_r^j V^{(0)})
\\ \nn &&\!\!\!\!\!\!\!\!\!\! 
- d_3^{\prime(1)} f_{abc} \int d^3{\bf x} \, \lim_{T_W \rightarrow \infty} 
g \lla G^a_{\mu\nu}({x}) G^b_{\mu\al}({x}) G^c_{\nu\al}({x}) \rra ,
\eea
\bea
&&\!\!\!\!\!\!\!\!\!\! 
V_{LS}^{(2,0)}(r) = {c_S^{(1)}\over 2 r^2}{\bf r}\cdot (\bfnabla_r V^{(0)})
\nn \\ && \!\!\!\!\!\!\!\!\!\! 
-{c_F^{(1)} \over r^2}i {\bf r}\cdot \lim_{T\rightarrow \infty}\int_0^{T}dt \, t \,  
\lla g{\bf B}_1(t) \times g{\bf E}_1 (0) \rra.  
\label{vls20}
\eea

For the $V^{(1,1)}$ potential we define
\begin{eqnarray*}
&&\!\!\!\!\!\!\!\!\!\! 
V^{(1,1)} = -{1 \over 2}\left\{{\bf p}_1\cdot {\bf p}_2,V_{{\bf p}^2}^{(1,1)}(r)\right\}
\\ &&\!\!\!\!\!\!\!\!\!\! 
-{V_{{\bf L}^2}^{(1,1)}(r)\over 2r^2}({\bf L}_1\cdot{\bf L}_2
+ {\bf L}_2\cdot{\bf L}_1)+ V_r^{(1,1)}(r)
\\ &&\!\!\!\!\!\!\!\!\!\! 
+ V_{L_1S_2}^{(1,1)}(r){\bf L}_1\cdot{\bf S}_2 - V_{L_2S_1}^{(1,1)}(r){\bf L}_2\cdot{\bf S}_1
\\ &&\!\!\!\!\!\!\!\!\!\! 
+ V_{S^2}^{(1,1)}(r){\bf S}_1\cdot{\bf S}_2 + V_{{\bf S}_{12}}^{(1,1)}(r){\bf S}_{12}({\hat {\bf r}}), 
\end{eqnarray*}
where ${\bf S}_{12}({\hat {\bf r}}) 
\equiv 3 {\hat {\bf r}}\cdot \bfsigma_1 \,{\hat {\bf r}}\cdot \bfsigma_2 
- \bfsigma_1\cdot \bfsigma_2$.  
Due to invariance under charge conjugation plus $m_1 \leftrightarrow m_2$ transformation, we have 
$V_{L_1S_2}^{(1,1)}(r)$ $=$ $V_{L_2S_1}^{(1,1)}(r; m_1 \leftrightarrow m_2)$.
Using Eqs. (\ref{singlet}) and (\ref{En2}) we get, in terms of Wilson loops,
\bea
&&\!\!\!\!\!\!\!\!\!\! \nn 
V_{{\bf p}^2}^{(1,1)}(r)=i{\hat {\bf r}}^i{\hat {\bf r}}^j
\lim_{T\rightarrow \infty}\int_0^{T}dt \, t^2 
\\  &&\quad \times 
\lla g{\bf E}_1^i(t) g{\bf E}_2^j(0) \rra_c,
\label{Em21p}
\eea
\bea
&&\!\!\!\!\!\!\!\!\!\! \nn 
V_{{\bf L}^2}^{(1,1)}(r)=i{\delta^{ij}-3{\hat {\bf r}}^i{\hat {\bf r}}^j \over 2}
\lim_{T\rightarrow \infty}\int_0^{T}dt \, t^2 
\\ &&\quad \times 
\lla g{\bf E}_1^i(t) g{\bf E}_2^j(0) \rra_c,
\label{Em21L}
\eea
\bea
&&\!\!\!\!\!\!\!\!\!\! 
V_r^{(1,1)}(r)= -{1 \over 2}(\bfnabla_r^2 V_{{\bf p}^2}^{(1,1)})
\label{Em21}
\\ \nn &&\!\!\!\!\!\!\!\!\!\! 
-i \lim_{T\rightarrow \infty}\int_0^{T}dt_1\int_0^{t_1} dt_2 \int_0^{t_2}
dt_3\, (t_2-t_3)^2 
\\ \nn &&\quad \times 
\lla g{\bf E}_1(t_1)\cdot g{\bf E}_1(t_2) g{\bf E}_2(t_3)\cdot g{\bf E}_2(0) \rra_c 
\\ \nn &&\!\!\!\!\!\!\!\!\!\! 
+ {1 \over 2} \Bigg(\bfnabla_r^i \lim_{T\rightarrow \infty}\int_0^{T}dt_1\int_0^{t_1} dt_2 (t_1-t_2)^2 
\\ \nn &&\quad \times 
\lla g{\bf E}_1^i(t_1) g{\bf E}_2(t_2)\cdot g{\bf E}_2(0) \rra_c  \Bigg)
\\ \nn &&\!\!\!\!\!\!\!\!\!\! 
+ {1 \over 2} \Bigg(\bfnabla_r^i \lim_{T\rightarrow \infty}\int_0^{T}dt_1\int_0^{t_1} dt_2 (t_1-t_2)^2 
\\ \nn &&\quad \times 
\lla g{\bf E}_2^i(t_1) g{\bf E}_1(t_2)\cdot g{\bf E}_1(0) \rra_c \Bigg)
\\ \nn &&\!\!\!\!\!\!\!\!\!\! 
- {i \over 2} \left(\bfnabla_r^i V^{(0)}\right)
\lim_{T\rightarrow \infty}\int_0^{T}dt_1\int_0^{t_1} dt_2  (t_1-t_2)^3 
\\ \nn &&\quad \times 
\lla g{\bf E}_1^i(t_1) g{\bf E}_2(t_2)\cdot g{\bf E}_2(0) \rra_c
\\ \nn &&\!\!\!\!\!\!\!\!\!\! 
- {i \over 2} \left(\bfnabla_r^i V^{(0)}\right)
\lim_{T\rightarrow \infty}\int_0^{T}dt_1\int_0^{t_1} dt_2  (t_1-t_2)^3 
\\ \nn &&\quad \times 
\lla g{\bf E}_2^i(t_1) g{\bf E}_1(t_2)\cdot g{\bf E}_1(0) \rra_c
\\ \nn &&\!\!\!\!\!\!\!\!\!\! 
- {1 \over 2} \lim_{T\rightarrow \infty}\int_0^{T}dt_1\int_0^{t_1} dt_2  (t_1-t_2)^2 
\\ \nn &&\quad \times 
\lla [{\bf D}_1.,g{\bf E}_1](t_1) g{\bf E}_2(t_2)\cdot g{\bf E}_2(0) \rra_c
\\ \nn &&\!\!\!\!\!\!\!\!\!\! 
+ {1 \over 2} \lim_{T\rightarrow \infty}\int_0^{T}dt_1\int_0^{t_1} dt_2  (t_1-t_2)^2 
\\ \nn &&\quad \times 
\lla [{\bf D}_2.,g{\bf E}_2](t_1) g{\bf E}_1(t_2)\cdot g{\bf E}_1(0) \rra_c
\\ \nn &&\!\!\!\!\!\!\!\!\!\! 
- {i \over 4} \lim_{T\rightarrow \infty}\int_0^{T}dt \, t^2
\lla [{\bf D}_1.,g{\bf E}_1](t) [{\bf D}_2.,g{\bf E}_2](0) \rra_c
\\ \nn &&\!\!\!\!\!\!\!\!\!\! 
+ {i \over 4} \Bigg(\bfnabla_r^i \lim_{T\rightarrow \infty}\int_0^{T}dt \, t^2
\Big\{ \lla g{\bf E}_1^i(t) [{\bf D}_2.,g{\bf E}_2](0) \rra_c
\\ \nn &&\quad  
- \lla g{\bf E}_2^i(t) [{\bf D}_1.,g{\bf E}_1](0) \rra_c \Big\} \Bigg)
\\ \nn &&\!\!\!\!\!\!\!\!\!\! 
- {1 \over 4} \lim_{T\rightarrow \infty}\int_0^{T}dt \, t^3
\Big\{ \lla [{\bf D}_1.,g{\bf E}_1](t) g{\bf E}_2^j (0) \rra_c 
\\ \nn &&\quad \times 
- \lla [{\bf D}_2.,g{\bf E}_2](t) g{\bf E}_1^j (0) \rra_c \Big\}
(\bfnabla_r^j V^{(0)})
\\ \nn &&\!\!\!\!\!\!\!\!\!\! 
+ {1 \over 4} \Bigg(\bfnabla_r^i
\lim_{T\rightarrow \infty}\int_0^{T}dt \, t^3 
\Big\{ \lla g{\bf E}_1^i(t) g{\bf E}_2^j (0) \rra_c 
\\ \nn &&\quad 
+ \lla g{\bf E}_2^i(t) g{\bf E}_1^j (0) \rra_c \Big\}  (\bfnabla_r^j V^{(0)}) \Bigg)
\\ \nn &&\!\!\!\!\!\!\!\!\!\! 
- {i \over 6} \lim_{T\rightarrow \infty}\int_0^{T}dt \, t^4
\\ \nn &&\quad \times 
\lla g{\bf E}_1^i(t) g{\bf E}_2^j (0) \rra_c
(\bfnabla_r^i V^{(0)}) (\bfnabla_r^j V^{(0)})
\\ \nn &&\!\!\!\!\!\!\!\!\!\! 
+ (d_{ss} + d_{vs} \lim_{T_W\rightarrow \infty}
\lla T_1^a T_2^{a} \rra ) \,\delta^{(3)}({\bf x}_1-{\bf x}_2),
\eea
\bea
&&\!\!\!\!\!\!\!\!\!\! \nn 
V_{L_2S_1}^{(1,1)}(r)= - {c_F^{(1)} \over r^2}i {\bf r}\cdot 
\lim_{T\rightarrow \infty}\int_0^{T}dt \, t \, 
\\ &&\quad \times  \lla g{\bf B}_1(t) \times g{\bf E}_2 (0) \rra, 
\label{vls11}
\eea
\bea
&&\!\!\!\!\!\!\!\!\!\! 
V_{S^2}^{(1,1)}(r) = {2 c_F^{(1)} c_F^{(2)} \over 3}i \lim_{T\rightarrow \infty}\int_0^{T} dt \,  
\lla g{\bf B}_1(t) \cdot g{\bf B}_2 (0) \rra \nn 
\\ &&\!\!\!\!\!\!\!\!\!\! 
- 4(d_{sv} + d_{vv} \lim_{T_W\rightarrow \infty}\lla T_1^a T_2^{a} \rra ) \,
\delta^{(3)}({\bf x}_1-{\bf x}_2), \label{vs2}
\eea
\bea
&&\!\!\!\!\!\!\!\!\!\! \label{vs12}
V_{{\bf S}_{12}}^{(1,1)}(r)=
{c_F^{(1)} c_F^{(2)} \over 4}i {\hat {\bf r}}^i{\hat {\bf r}}^j  \lim_{T\rightarrow \infty}\int_0^{T} dt \, 
\\ &&\!\!\!\!\!\!\!\!\!\! 
\times \Big[ \lla g {\bf B}^i_1(t) \, g {\bf B}^j_2 (0) \rra  
- {\delta^{ij}\over 3}\lla g{\bf B}_1(t) \cdot g{\bf B}_2 (0) \rra \Big].
\nn
\eea

A critical comparison of the above results with the previous literature will be done in the conclusions. 
Here, I mention that the potentials (\ref{Em12}), (\ref{Em20}) and (\ref{Em21})  
have been calculated for the first time in Ref. \cite{m1} and \cite{m2}. 
Since the potential we get here is a well defined quantity, derived from QCD via a systematic 
and unambiguous procedure, and {\it complete} up to order $1/m^2$, 
it is not affected by the usual ambiguities (ordering, retardation corrections, etc.), which affect 
all potential models and all phenomenological reductions of Bethe--Salpeter kernels \cite{rev0}. 
For the same reason the above result may be relevant for the study of the properties 
of the QCD vacuum in the presence of heavy sources. While the lattice data for the spin structure of the 
potential seem to suggest the so-called ``scalar confinement'', the data for the momentum 
dependent and spin independent part of the potential seem to support 
a flux-tube picture \cite{rev0,flux}. It will be interesting to see how consistent 
these pictures are with the new momentum and spin independent potentials, once lattice 
data will be available for them. I note that some of them are not simply expressed by two field 
insertions on a static Wilson loop. In particular an extended object coming from the Yang--Mills 
sector is required (similar extended objects would also show up by taking into account operators 
with light quarks). 


\section{POWER COUNTING}
\label{secpc}
The standard power counting of NRQCD, as discussed, for instance, in \cite{pc},
has been tested in the perturbative regime. However, even in this regime, due 
to the different dynamical scales still involved, the matrix elements of NRQCD 
do not have an unique power counting in $v$. In the non-perturbative regime 
the problem of the power-counting of NRQCD is still open. The above non-perturbative 
formulation of pNRQCD has translated this problem to obtaining 
the power counting of the different potentials. This is expected to be of some 
advantage. Since the scale $mv$ has been integrated out, 
the power counting of pNRQCD is simpler. Since all the potentials are expressed 
in terms of Wilson loops, there are or there will  be direct lattice measurements of them.  

Here, let us only say few words about the expected behaviour of the potential using 
arguments of naturalness on the scale $mv$, i.e. assuming that the potentials scale with $mv$.  
We first consider $V^{(0)}$. In principle, $V^{(0)}$ counts as $mv$, but, 
by definition, the kinetic energy counts as $mv^2$. Therefore, 
the virial theorem constrains $V^{(0)}$ also to count as $mv^2$. 
In the perturbative case this extra $v$ suppression comes from the factor $\als \sim v$ in the potential. 
Using naturalness, $V^{(1,0)}/m$ scales like $mv^2$. Therefore, it could be in principle as
large as $V^{(0)}$. This makes a lattice calculation of this potential urgent.
Perturbatively, due to the factor $\als^2$, it is $O(mv^4)$.
For what concerns the $1/m^2$ potentials, the naturalness argument suggests that they are of order $mv^3$. 
However, also here several constraints apply. Terms involving $\bfnabla V^{(0)}$ are suppressed 
by an extra factor $v$, due to the virial theorem. The Gromes relations \cite{spin2}, 
$\displaystyle {1\over 2r}{d V^{(0)}\over dr}$  $+$  $V_{LS}^{(2,0)}$ 
$-$  $V_{L_2S_1}^{(1,1)}$  $=$  $0$, 
suppresses by an extra factor $v$ the combination $V_{LS}^{(2,0)}$ $-$ $V_{L_2S_1}^{(1,1)}$. 
Similar constraints also exist for the spin-independent potentials \cite{BMP}.
Finally, it may be important to consider that some of the potentials are $O(\als)$ suppressed 
in the matching coefficients inherited from NRQCD.

\section{CONCLUSIONS AND OUTLOOK}
\label{conclusions}
I have reported about a new derivation of the QCD potential in an effective field theory framework. 
Explicit expressions for the heavy quarkonium potential up to order $1/m^2$, 
valid beyond perturbation theory, have been written.

I now compare with previous results in the literature. 
For the spin-dependent potentials we find agreement with the Eichten--Feinberg results
\cite{spin1,spin2} (once the NRQCD matching coefficients have been taken into account) {\it except} 
for the $1/m_1m_2$ spin-orbit potential $V_{L_2S_1}^{(1,1)}$. 
Our result gives back the well-known tree-level calculation, whereas the Eichten--Feinberg
expression gives half of the expected result. Moreover, our perturbative result fulfills the  Gromes 
relation. An analysis of this error, present in the original papers and 
in several others appeared afterwards, is done in \cite{m2}. 
The spin-independent potentials have been computed before only by Barchielli, Brambilla, 
Montaldi and Prosperi in \cite{BMP} (the analysis done in \cite{thesis}, which appears 
to be inconclusive, has never been published). We agree (once the NRQCD matching coefficients
have been taken into account) with their result for the momentum-dependent
terms but not for the momentum-independent terms, where new contributions are found. 
An approach similar to that one presented here has been used in \cite{SS} 
in order to derive, from the QCD Hamiltonian in Coulomb gauge, the spin-dependent part of the
potential up to $O(1/m^2)$. Our expression (\ref{En2}) differs from that one used 
in \cite{SS}, which, in general, would give incorrect spin-independent potentials. 
However, if we take our matching coefficients at tree level and neglect the tree-level 
annihilation contributions in the equal mass case, we find essential agreement for the spin-dependent 
potentials.  

I conclude, commenting on two possible developments of the present work.
First, it is worthwhile to explore the possibility of expressing the potentials 
associated with higher gluonic excitations in terms of Wilson loop operators as done here 
for the heavy quarkonium ground state. The corresponding quantum-mechanical expressions  
are known and have been calculated in \cite{m2}.
Second, our results are complete at $O(1/m^2)$ in the case of pure gluodynamics. If we
want to incorporate light fermions, the procedure to follow is analogous and our results still remain 
valid (considering now matrix elements and Wilson loops with dynamical light fermions incorporated) 
except for new terms appearing in the energies at $O(1/m^2)$ due to operators involving light fermions. 
They may be incorporated along the same lines as the terms discussed here.

{\bf Acknowledgments}

I thank Nora Brambilla, Joan Soto and especially Antonio Pineda 
for collaboration on the work presented here.
I thank Nora Brambilla for reading the manuscript and comments.  
I thank the organizers for invitation and support and the Alexander von Humboldt foundation  
for support.

1. {\bf V. Zakharov (MPI, Munich)}: {\it At this conference a very big 
 perturbative correction was reported by Penin at the next order, 
 which has not been explored earlier. Does it affect in any way your 
 approach or result?}

{\bf A. Vairo}: {\it The correction mentioned by Alexander Penin was first derived in \cite{pnrqcd1}. 
 It arises from the leading US corrections (i.e. from the NLO in the multipole expansion) to the 
 pNRQCD Lagrangian in the perturbative regime.}

2. {\bf G. Martinelli (Univ. of Rome and Orsay)}: {\it May you comment on the implementation 
 of NRQCD in lattice calculations?}

{\bf A. Vairo}: {\it The implementation of NRQCD on the lattice depends crucially on the 
 power counting in the non-perturbative regime. Once the power counting in $v$ is established, 
 the matching coefficients, which are counted in $\als$, have to be included consistently.  
 Due to the breaking of Lorentz invariance on the lattice, new terms, with respect 
 to those mentioned here, may arise. \vspace{-2mm}}

\end{document}